\documentclass[3p,sort&compress]{elsarticle}

\usepackage[english]{babel} 
\usepackage[T1]{fontenc}
\usepackage[utf8x]{inputenc}

\usepackage{url}

\usepackage[boxed]{algorithm2e}
\usepackage{algorithmic}
\usepackage{amsmath}
\usepackage{amssymb}
\usepackage{subcaption}
\usepackage[dvipsnames]{xcolor}
\usepackage{pgf}
\usepackage{capt-of}
\usepackage{mathrsfs}

\usepackage[pdftex]{hyperref} 
\hypersetup{colorlinks=true}
\hypersetup{linkcolor=black}
\hypersetup{citecolor=black}

\usepackage{tikz}
\usepackage{pgfplots}
\pgfplotsset{compat=1.5.1}
\usetikzlibrary{calc}
\usetikzlibrary{arrows}

\newcommand{\bs}[1] {\boldsymbol{#1}}
\newcommand{\fp}{{f\rightarrow p}}
\newcommand{\pf}{{p\rightarrow f}}
\newcommand{\eps}{\varepsilon}

\begin{document}
	
\title{A Coupled Lattice Boltzmann Method and Discrete Element Method for Discrete Particle Simulations of Particulate Flows}

\author[lss]{C. Rettinger\corref{cor1}}
\author[lss,cerfacs]{U. R\"{u}de}

\address[lss]{Chair for System Simulation, Friedrich--Alexander--Universität Erlangen--Nürnberg, Cauerstraße 11, 91058 Erlangen, Germany}
\address[cerfacs]{CERFACS, 42 Avenue Gaspard Coriolis, 31057 Toulouse Cedex 1, France}

\cortext[cor1]{E-mail: christoph.rettinger@fau.de}

\begin{abstract}
Discrete particle simulations are widely used to study large--scale particulate flows in complex geometries where particle--particle and particle--fluid interactions require an adequate representation but the computational cost has to be kept low.
In this work, we present a novel coupling approach for such simulations.
A lattice Boltzmann formulation of the generalized Navier--Stokes equations is used to describe the fluid motion. 
This promises efficient simulations suitable for high performance computing and, since volume displacement effects by the solid phase are considered, our approach is also applicable to non--dilute particulate systems.
The discrete element method is combined with an explicit evaluation of interparticle lubrication forces to simulate the motion of individual submerged particles.
Drag, pressure and added mass forces determine the momentum transfer by fluid--particle interactions.
A stable coupling algorithm is presented and discussed in detail.
We demonstrate the validity of our approach for dilute as well as dense systems by predicting the settling velocity of spheres over a broad range of solid volume fractions in good agreement with semi--empirical correlations.
Additionally, the accuracy of particle--wall interactions in a viscous fluid is thoroughly tested and established.
Our approach can thus be readily used for various particulate systems and can be extended straightforward to e.g. non--spherical particles.
\end{abstract}


\maketitle
	
\section{Introduction}

In the broad field of particulate flows, simulations are becoming increasingly important to study their complex behavior and to enhance our understanding of the underlying processes.
Ideally, one would like to apply so--called direct numerical simulations (DNS) that are characterized by geometrically fully resolved particles and a fine enough numerical resolution to capture all scales of fluid motion.
They have been applied successfully to e.g. fluidized beds \cite{gotz_direct_2010, xiong_large-scale_2012,Esteghamatian17} and sediment beds in rivers \cite{KIDANEMARIAM2014174,rettinger2017fully,kidanemariam_uhlmann_2017,Seil2017}.
As a main benefit, they enable a detailed analysis of flow properties inside closely packed particle clusters and naturally allow for an accurate evaluation of trajectories and interaction forces of individual particles.
Such quantities are usually difficult to obtain in laboratory experiments.
However, the large number of particles typically encountered in such systems poses a major challenge for DNS and restricts their applicability to relatively small system sizes with currently at most $\mathcal{O}(10^6)$ particles \cite{gotz_direct_2010,rettinger2017fully,kidanemariam_uhlmann_2017}.

Alternative simulation approaches exist that introduce certain degrees of modeling to account for then unresolved physical processes.
This decreases the computational cost significantly and enables simulations of larger systems over a longer period of time.
Those approaches can be roughly classified as either two--fluid models (TFM) or discrete particle simulation models (DPS) \cite{van_der_hoef_2008}.
The TFM is an Eulerian--Eulerian model where the solid phase, like the fluid phase, is regarded as a continuum and its evolution is described by a set of additional model equations.
This approach requires that several modeling assumptions are made and naturally can not represent the details of particle--particle or particle--fluid interactions adequately \cite{van_der_hoef_2008}.
In DPS models, the solid phase is represented by the actual particles, using a Lagrangian description of their motion.
Particle collisions are usually treated with the discrete element method (DEM) \cite{cundall1979discrete}.
Different from a DNS, however, the particle size is typically of the order or smaller than the numerical grid spacing used for the representation of the fluid phase. 
The fluid--particle interaction, as well as possible turbulent flow structures, are thus no longer resolved by the numerical simulation and have to be modeled.
The overall idea of these approaches is to establish the coupling of the fluid and the solid phase via interaction forces.
For each particle, such a fluid--particle interaction force is evaluated and applied, and a corresponding reaction force is employed on the surrounding fluid.
Therefore, closure correlations for most of the different contributions to this interaction force are required.
In most applications, the drag force is the most important contribution and several correlations have been proposed here \cite{BOGNER201571,tang_2015}.

DPS is a well--established method in classical CFD, where it is often referred to as CFD--DEM.
It is also the basis of several commercial tools and frequently used in various application fields, see the reviews in \cite{SUBRAMANIAM2013215,ZHONG201616,gediscrete}.
In recent years, the lattice Boltzmann method (LBM) has become a prominent alternative to those CFD methods \cite{chen_lattice_1998}.
Since it essentially requires only local operations to simulate fluid flow, it is well--suited for highly parallel and scalable simulations to be executed an large supercomputers \cite{godenschwager_framework_2013}. 
This feature makes the LBM a viable choice for DNS \cite{gotz_direct_2010, xiong_large-scale_2012, rettinger2017fully,Seil2017}.
Furthermore, different approaches have been developed to carry out DPS by coupling the LBM with the DEM \cite{Ahlrichs_1999, ZHANG20151, Xiong2014, WANG2013228}.
We will briefly review these LBM--DEM DPS approaches and summarize their properties and areas of applicability, see Sec.~\ref{sec:review}. 
As we will see, there is currently no single approach available that has demonstrated to yield satisfactory results for the whole range of dilute to dense particulate systems in terms of accuracy and stability.

The aim of this work is thus to develop a new coupled LBM--DEM scheme for DPS that is carefully validated and can then be readily applied to different scenarios.
It should retain the unique features of the LBM that are important for the efficient parallel execution and allow for a flexible extension of the applied models if required by a specific simulation.
Additionally, measures for a stable momentum transfer between the phases have to be considered.
We will mainly focus on fluid--solid systems with spherical particles which will guide our choice of interaction force contributions, namely lubrication and added mass forces, and the empirical correlations that we include in our model.
Nevertheless, the resulting approach will also be applicable to gas--solid flows and can be extended to non--spherical particles.

The remainder of this work is structured as follows: 
After the review of existing LBM--DEM DPS approaches in Sec.~\ref{sec:review}, we present our numerical model in detail in Sec.~\ref{sec:method}.
This is split into discussions of the treatment of the solid phase in Sec.~\ref{sec:particle_phase} and the fluid phase in Sec.~\ref{sec:fluid_phase}.
The complete algorithm is summarized in Sec.~\ref{sec:DPSalgo}.
We demonstrate the validity of our approach for the prediction of settling velocities over a large range of solid volume fractions in Sec.~\ref{sec:hinderedSettling}.
The accuracy of particle--wall interactions is analyzed in Sec.~\ref{sec:sphere_wall_collision}.
In Sec.~\ref{sec:conclusion}, we summarize the approach and the results, and give an outlook to further developments.

\section{Review of existing LBM--DEM DPS approaches}
\label{sec:review}

Several DPS approaches that couple the LBM and the DEM have been proposed.
We briefly review their main features and discuss their applicability.
We subdivide the DPS approaches into two variants: the two--way coupling and the four--way coupling approaches.
They all have in common that interaction forces between the particles and the fluid phase are evaluated, typically via interpolations to transfer information from the Eulerian (fluid) to the Lagrangian (particle) description and vice versa.
This force is then applied onto the particles.
A corresponding reaction force is employed on the surrounding fluid cells.
In this overview, we do not consider approaches that neglect the effect of the particles onto the fluid, referred to as one--way coupling.

\subsection{Two--way coupling approaches}
In these approaches, the fluid flow is described via the incompressible Navier--Stokes equations and the interaction between the two phases is established solely via interaction forces.
In the context of the LBM, this was explored in a pioneering paper \cite{Ahlrichs_1999} for the coupling with molecular dynamics simulations.
Here, the interaction force $\bs{F}_{\fp} = - \zeta ( \bs{u}_p - \bs{u}_f(\bs{x}_p,t) )$, with a friction coefficient $\zeta$, is evaluated at the polymer position $\bs{x}_p$ and requires the interpolated fluid velocity $\bs{u}_f$ as well as the polymer velocity $\bs{u}_p$. 
An overview of this method as well as further developments can be found in \cite{Duenweg2009}.

Another method is the so--called \textit{Particulate Immersed Boundary Method} (PIBM) from \cite{ZHANG20151} that can be interpreted as a limiting case of the immersed boundary method of \cite{NIU2006173}. 
The interaction force is calculated by interpolating the particle distribution functions available in the LBM, see Sec.~\ref{sec:LBM}, to the particle position and applying the momentum exchange formulation in LBM.
It can be shown that this interaction force density takes the form $\bs{f}_{\fp,\text{PIBM}} = 2\rho_f A_p( \bs{u}_f(\bs{x}_p,t) - \bs{u}_p)$, see e.g. \cite{HU2014140}, with $A_p$ the cross sectional area of the particle.
Thus, the PIBM is very similar to the aforementioned coupling approach of \cite{Ahlrichs_1999}.
It has certain inconsistencies in terms of the physical units of $\bs{f}_{\fp,\text{PIBM}}$, and it lacks the contribution from the fluid viscosity in comparison to e.g. Stokes' drag force.
This might explain the undesired viscosity dependency of the single sphere settling velocity, as observed in \cite{ZHANG20151}.
Recently, the PIBM approach has been enhanced by modifying the PIBM force with empirical drag correlations \cite{HABTE2017486}, which has been further investigated by \cite{CEAT:CEAT201600547}.
Both studies confirm that the enhanced PIBM is capable of predicting the settling velocity of a single sphere correctly, but a thorough investigation of its accuracy for non--dilute systems is still missing.

\subsection{Four--way coupling approaches}
As has been investigated in \cite{cihonski_finn_apte_2013}, it is crucial to account for fluid volume displacement effects when considering non--dilute systems, or particles with sizes comparable to the size of the control volumes, i.e. grid cells, to obtain accurate results.
Such approaches are often referred to as \textit{four--way} or \textit{volumetric} coupling approaches since they typically include the local fluid volume fraction $\eps_f$ into the Navier Stokes equations.
A commonly used variant are the volume--averaged Navier Stokes (VANS) equations \cite{JOSEPH199035}, especially in the area of gas--fluidized beds \cite{KUIPERS19921913,van_der_hoef_2008}.
A LBM formulation of the VANS equations exists \cite{wang_wang_2005,Zhang2014} and has been combined with empirical drag correlations to carry out simulations of gas--solid fluidized beds \cite{Xiong2014}.
However, it has been shown that this formulation yields spurious velocities in the presence of larger solid volume fraction gradients which could render the simulations inaccurate and unstable in heterogeneous systems \cite{BLAIS2015258}.
The proposed new scheme in \cite{BLAIS2015258} overcomes these problems but introduces several correction terms that in turn render the LBM inefficient and tedious.
An alternative LBM formulation of the VANS equations exists \cite{SONG2013442} which, however, necessitates the inclusion of mass sources and a local solution of matrix systems, which again adds substantial complexity to the LBM.

Another variant of such a coupling approach was presented in \cite{WANG2013228}, which uses the partially saturated cells method (PSM) from \cite{noble_lattice-boltzmann_1998} to include the effect of fluid volume displacement.
This method is usually applied in fully resolved simulations, as e.g. in \cite{RETTINGER201774} where particles span several grid cells.
Since the particles in DPS are smaller than a grid cell, the computed forces from PSM are expected to be inaccurate and are thus rescaled in \cite{WANG2013228} with the help of empirical drag laws.
This approach has been applied successfully to gas--solid fluidization simulations.
However, a thorough analysis of the PSM is not yet available and it thus remains unclear which macroscopic equations are solved with the approach.
Additionally, the applicability of this approach for dilute systems is questionable since the proposed rescaling relies on averaging the interaction force over all particles inside a computational cell and thus the accuracy might deteriorate if only a few particles are present in a cell.

\section{Numerical Method}

\label{sec:method}

In this section, we present the equations describing the particle motion, Sec.~\ref{sec:particle_phase}, and the fluid dynamics, Sec.~\ref{sec:fluid_phase}.
We discuss the applied numerical methods and the included models to account for various physical effects.
All these components are implemented in the open--source LBM framework \textsc{waLBerla}\footnote{\url{http://walberla.net}}.
A summary of the complete algorithm is given in Sec.~\ref{sec:DPSalgo}.

\subsection{Solid phase}
\label{sec:particle_phase}

The motion of a particle, given by the temporal evolution of its position $\bs{x}_p$, linear velocity $\bs{u}_p$ and angular velocity $\bs{\omega}_p$, is described by the following set of equations
\begin{subequations}
	\label{eq:particle_motion}
	\begin{align}
	\frac{\text{d}\bs{x}_p}{\text{d} t} & = \bs{u}_p, \\
	\rho_p V_p \frac{\text{d}\bs{u}_p}{\text{d} t} & = \bs{F}_{c} + \bs{F}_\fp + \bs{F}_{lub} + \bs{F}_g, \\
	I_p \frac{\text{d} \bs{\omega}_p}{\text{d} t} & = \bs{T}_{c},
	\end{align}
\end{subequations}
with the particle density $\rho_p$, volume $V_p$ and moment of inertia $I_p$.
The net force acting on a particle consists of different contributions.
These are: collision forces $\bs{F}_c$, fluid--particle interaction forces $\bs{F}_\fp$, lubrication forces $\bs{F}_{lub}$, and gravity $\bs{F}_g$.
The fluid--particle interaction force is the sum of all interphase momentum transfer contributions which are discussed in Sec.~\ref{sec:particle_Ffp}.
The time integration of above equations is carried out by an explicit time--stepping scheme \cite{Preclik2015} with a time step $\Delta t_{\text{DEM}}$ which is efficient and accurate enough for the regarded systems \cite{van_der_hoef_2008}.

\subsubsection{Discrete element method}
\label{sec:dem}
\begin{figure}[t]
\centering
\begin{tikzpicture}[]
\coordinate[label=left:$\bs{x}_{p,i}$] (xpi) at (2,2);
\coordinate[label=right:$\bs{x}_{p,j}$] (xpj) at (5.4,2.5);
\coordinate[label=above:$C$] (C) at ($ (xpj)!0.5!(xpi) $ );
\coordinate (nij) at ($0.5*(xpj)-0.5*(C)$ );
\coordinate (tij) at ([rotate=-90]nij);

\draw[gray, <->] ($(xpi)+(0,-2)$) -- ++(0,4) node[pos=0.8,left]{$d_{p,i}$};
\draw[gray, <->] ($(xpj)+(0,-2)$) -- ++(0,4) node[pos=0.8,right]{$d_{p,j}$};
\node[fill=black, circle, inner sep=1.5] at (xpi) {};
\node[fill=black, circle, inner sep=1.5] at (xpj) {};
\node[fill=black, circle, inner sep=1.5] at (C) {};
\draw (xpi) circle (2);
\draw (xpj) circle (2);
\draw[dashed] (xpi) -- (xpj);
\draw[-latex, thick] (C) -- ++(nij) node[pos=1,below]{$\bs{n}_{ij}$};
\draw[-latex, thick] (C) -- ++(tij) node[pos=1,right]{$\bs{t}_{ij}$};
\draw[-latex, thick] (xpi) -- ++(1.8,-1.5) node[pos=1,right]{$\bs{u}_{p,i}$};
\draw[-latex, thick] (xpj) -- ++(-0.4,1.3) node[pos=1,left]{$\bs{u}_{p,j}$};

\coordinate (intersect1) at ($(C)-0.33*(nij)$);
\coordinate (intersect2) at ($(C)+0.33*(nij)$); 
\draw[blue] (intersect1) -- ++($-2.7*(tij)$);
\draw[blue] (intersect2) -- ++($-2.7*(tij)$);
\draw[thick,<->,blue] ($(intersect1)-2.5*(tij)$) -- ($(intersect2)-2.5*(tij)$) node[pos=0.5,above]{$\delta_{ij}$};

\draw[-latex,thick] ($(xpi) + (85:2.3)$) arc (85:150:2.3);
\node[left] at ($(xpi) + (150:2.3)$) {$\bs{\omega}_{p,i}$};
\draw[-latex,thick] ($(xpj) + (-10:2.3)$) arc (-10:75:2.3);
\node[above] at ($(xpj) + (75:2.3)$) {$\bs{\omega}_{p,j}$};
\end{tikzpicture}
\caption{Schematic representation of two colliding spheres, $i$ and $j$, together with reference quantities required by the DEM collision model.}
\label{fig:DEM}
\end{figure}
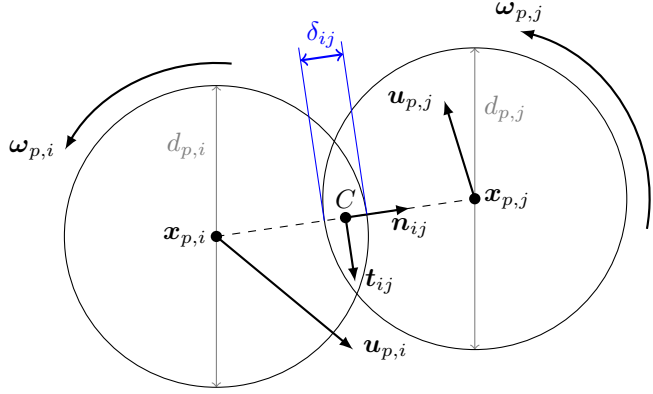

To compute the forces and torques arising from particle--particle or particle--wall collisions, the discrete element method (DEM) from \cite{cundall1979discrete} is used.
Collisions are modeled based on a linear mass--spring--damper system and thus result in a so--called soft--sphere model.
In this work, we will restrict ourselves to spherical particles for simplicity, noting however that an extension to more complex shapes is possible \cite{Iglberger2010}. 
When considering two particles, denoted by indices $i$ and $j$, inside a system of $N_p$ particles, we can define an overlap length 
\begin{equation}
\delta_{ij} = \tfrac{1}{2}(d_{p,i} + d_{p,j}) - |\bs{x}_{p,j}-\bs{x}_{p,i}|,
\end{equation}
with the particle diameter $d_p$ and its center position $\bs{x}_p$, see the sketch in Fig.~\ref{fig:DEM}.
The contact normal unit vector of the collision is defined as 
\begin{equation}
\label{eq:collision_normal}
\bs{n}_{ij} = \frac{\bs{x}_{p,j}-\bs{x}_{p,i}}{|\bs{x}_{p,j}-\bs{x}_{p,i}|},
\end{equation}
and the relative velocity at the contact point $C$ is
\begin{equation}
\bs{u}_{r,ij} = \bs{u}_{p,i} - \bs{u}_{p,j} + (\tfrac{1}{2}d_{p,i}\bs{\omega}_{p,i} + \tfrac{1}{2}d_{p,j}\bs{\omega}_{p,j})\times\bs{n}_{ij}.
\end{equation}
This velocity can be decomposed into the normal and the tangential part, given as
\begin{subequations}
	\begin{align}
	\bs{u}_{r,ij}^n &= (\bs{u}_{r,ij} \cdot \bs{n}_{ij} ) \bs{n}_{ij},\\
	\bs{u}_{r,ij}^t &= \bs{u}_{r,ij} - \bs{u}_{r,ij}^n,
	\end{align}
\end{subequations}
respectively.

Following the spring--damper collision model, with a stiffness constant $k_n$ and a normal damping constant $\xi_n$, the normal collision force on particle $i$ due to this contact is
\begin{equation}
\bs{F}_{c,ij}^n = 
\begin{cases}
- k_n \delta_{ij}\bs{n}_{ij} - \xi_n \bs{u}_{r,ij}^n, &\delta_{ij} > 0, \\
0, &\text{else.}
\end{cases}
\end{equation} 
The collision force in tangential direction distinguishes between a sliding and a sticking regime, and is given as
\begin{equation}
\bs{F}_{c,ij}^t = 
\begin{cases}
- \text{min}(\mu_c |\bs{F}_{c,ij}^n|, \xi_t|\bs{u}_{r,ij}^t|) \bs{t}_{ij}, & \text{if } |\bs{u}_{r,ij}^t| > 0, \\
0, & \text{else,}
\end{cases}
\end{equation}
with the tangential unit vector $\bs{t}_{ij} =  \bs{u}_{r,ij}^t/|\bs{u}_{r,ij}^t|$, the friction coefficient $\mu_c$, and the tangential damping constant $\xi_t$.
Consequently, the total force on particle $i$ due to possible collisions with all other particles in the system is then given as
\begin{equation}
\label{eq:forces_collision}
\bs{F}_c = \sum_{j=1,i\neq j}^{N_p} (\bs{F}_{c,ij}^n + \bs{F}_{c,ij}^t).
\end{equation}

The tangential collision force also generates a torque on the particle and thus the total collisional torque on particle $i$ is
\begin{equation}
\label{eq:torques_collision}
\bs{T}_c = \sum_{j=1,i\neq j}^{N_p} \frac{d_{p,i}}{2} \bs{F}_{c,ij}^t \times \bs{n}_{ij}.
\end{equation}

In total, this DEM model introduces four parameters, namely $k_n, \xi_n, \xi_t$ and $\mu_c$.
In our simulations, we always choose $\xi_n = \xi_t$.
Additionally, we use
\begin{equation}
\label{eq:damping_to_stiffness}
	\xi_n = \frac{-2 \ln e_n \sqrt{m_{ij} k_n}}{\sqrt{\pi^2+\ln^2 e_n}},
\end{equation}
which is an analytical solution of the linear spring--damper problem \cite{crowe2011multiphase}, and relates $\xi_n$ to $k_n$.
It features the dry coefficient of restitution, $e_n$, that is often available in literature for different materials, and the reduced mass $m_{ij}= m_{p,i}m_{p,j}/(m_{p,i} + m_{p,j})$ for a particle pair and $m_{ij} = m_{p,i}$ for a particle--wall contact.
Furthermore, the duration of a collision event can be estimated as \cite{crowe2011multiphase}:
\begin{equation}
\label{eq:collision_time}
T_c = \frac{2\pi m_{ij}}{\sqrt{4 m_{ij} k_n - \xi_n^2}}.
\end{equation}
By choosing a specific $T_c$, and together with Eq.~\eqref{eq:damping_to_stiffness}, the stiffness and damping coefficients can be computed.

\subsubsection{Fluid--particle interaction forces}
\label{sec:particle_Ffp}
\begin{table}[t]
	\centering
	\begin{tabular}{ll}
		force & formula \\[3pt] \hline
		drag force & $\bs{F}_{d} = 3\pi d_p \mu_0 (1-\eps_{p|p}) C_d(Re_{p|p}, \eps_{p|p}) ( \bs{u}_{f|p} - \bs{u}_p )$, \\[3pt]
		& $C_d(Re_p,\eps_p) = ( 1-\eps_p)\left(\frac{C_d(Re_p,0)}{(1-\eps_p)^3} + A(\eps_p) + B(Re_p,\eps_p) \right)$ \\[5pt]
		& $C_d(Re_p,0) = 1 + 0.15 Re_p^{0.687}$ \\[3pt]
		& $A(\eps_p) = \frac{5.81\eps_p}{(1-\eps_p)^3} + 0.48 \frac{\eps_p^{1/3}}{(1-\eps_p)^4}$ \\[3pt]
		& $B(Re_p,\eps_p) = \eps_p^3 Re_p\left(0.95+\frac{0.61\eps_p^3}{(1-\eps_p)^2}\right)$ \\[3pt]
		pressure gradient force &  $\bs{F}_{pr} = - V_p (\nabla P)_{|p}$ \\[3pt]
		lift force & $ \bs{F}_{l} = 1.61 d_p^2 \sqrt{\frac{\mu_0 \rho_f}{|\nabla \times \bs{u}_f|_{|p}}} \left( (\bs{u}_{f|p}-\bs{u}_p) \times ( \nabla \times \bs{u}_f )_{|p} \right)$ \\[3pt]
		added (virtual) mass force & $\bs{F}_{am} = C_{am} \rho_f V_p \left( \left(\frac{\text{D} \bs{u}_f}{\text{D}t}\right)_{|p} - \frac{\text{d}\bs{u}_p}{\text{d}t}\right), C_{am} = 0.5$ \\
		 \hline
	\end{tabular}
	\caption{Considered contributions to the fluid--particle interaction force and the applied (closure) relations.}
	\label{tab:InteractionForces}
\end{table}
The hydrodynamic force acting on the individual particles contains several contributions that account for different physical effects.
Specifically, we here include drag $\bs{F}_d$, pressure $\bs{F}_{pr}$, lift $\bs{F}_l$, and added mass $\bs{F}_{am}$ effects:
\begin{equation}
\label{eq:forces_fp}
\bs{F}_\fp = \bs{F}_d + \bs{F}_{pr} + \bs{F}_l + \bs{F}_{am}
\end{equation}
The choice for or against the inclusion of certain effects depends on the targeted setup.
In simulations of gas--solid systems, such as gas--fluidized beds, often only the first two components are modeled.
In fluid--solid systems, like sediment transport in river beds, the lift and added mass contributions cannot be neglected and are thus typically included \cite{finn_li_apte_2016,SUN2016207}.
The models for the different contributions are displayed in Tab.~\ref{tab:InteractionForces} and are briefly discussed next.

\begin{figure}[t]
\centering
\begin{tikzpicture}[]
\coordinate (xp) at (3.2,1.9);
\node[circle,fill=lightgray,draw,minimum size=0.8cm] (part) at (xp){};
\node[circle,fill=black,inner sep=1.5] at (xp){};
\node[below] at (xp){$\bs{x}_{p}$};
\node[circle,draw,minimum size=3cm,blue] at (xp){};
\draw (0,0) grid (7,4);
\foreach \x in {1,...,7}{
	\foreach \y in {1,...,4}{
		\node[fill=black,inner sep=1.5pt] at ($(\x,\y) - (0.5,0.5)$){}; 
	} 
}
\draw[thick] (3.5,0.5) -- (part);
\draw[thick] (2.5,1.5) -- (part);
\draw[thick] (3.5,1.5) -- (part);
\draw[thick] (4.5,1.5) -- (part);
\draw[thick] (2.5,2.5) -- (part);
\draw[thick] (3.5,2.5) -- (part);
\draw[thick] (4.5,2.5) -- (part);

\draw (0,0) -- (0,-0.2);
\draw (1,0) -- (1,-0.2);
\draw[thick,<->] (0,-0.15) -- (1,-0.15) node[pos=0.5,below]{$\Delta x$};

\draw (0,0) -- (-0.2,0);
\draw (0,1) -- (-0.2,1);
\draw[thick,<->] (-0.15,0) -- (-0.15,1) node[pos=0.5,left]{$\Delta x$};
\end{tikzpicture}
\caption{Schematic representation of the interpolation operator in Eq.~\eqref{eq:filter_def}. The gray circle depicts a particle at position $\bs{x}_p$ on an Eulerian grid, with cell centers $\bs{x}$ marked by black squares. The blue circle around the particle is of radius $1.5\Delta x$ and thus shows the extent of the filter kernel. All cells with cell centers inside this circle contribute to the interpolation, displayed by the black lines.}
\label{fig:Interpolation}
\end{figure}
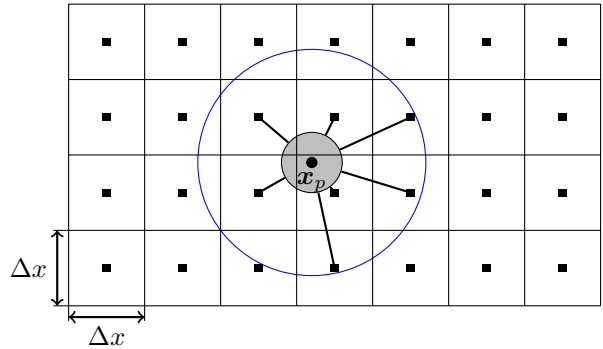

In all cases, different Eulerian field variables must be evaluated at the particle position, which will be denoted as $\phi_{|p}$. 
This requires an interpolation from the cell centers $\bs{x}$ of the Eulerian grid cells to the particle position $\bs{x}_p$ and is given as an interpolation operator $\mathscr{I}(\bs{x},\bs{x}_p)$, such that:
\begin{equation}
\label{eq:interpolation}
\phi_{|p} = \sum_{\bs{x}} \mathscr{I}(\bs{x},\bs{x}_p) \phi
\end{equation}
In this work, we use the three point discrete delta functions from \cite{ROMA1999509} that we will also need for the opposite operation later in Sec.~\ref{sec:GNS}:
\begin{equation}
\label{eq:filter_def}
\mathscr{I}(\bs{x},\bs{x}_p) = 
\begin{cases}
\tfrac{1}{3}(1+\sqrt{-3r^2+1}, & r < 0.5, \\
\tfrac{1}{6}\left(5-3r-\sqrt{-3(1-r)^2+1}\right), & 0.5 \leq r < 1.5, \\ 
0, & 1.5 \leq r,
\end{cases}
\end{equation}
with $r = |\bs{x}-\bs{x}_p|/\Delta x$ and the Eulerian grid spacing $\Delta x$, see also Fig.~\ref{fig:Interpolation}.
Other possibilities are trilinear interpolation \cite{finn_li_apte_2016}, Lagrange polynomial interpolation \cite{ZHANG20151}, or other filter kernels \cite{YANG20097821}.
Note, that apart from the fluid quantities like velocity and pressure, we also interpolate the solid volume fraction $\eps_p$.

The drag force acting on a single sphere in the Stokes flow regime is given by Stokes' law as $\bs{F}_{d,St} = 3 \pi d_p \mu_0 (\bs{u}_{f|p} - \bs{u}_p)$.
In order to extend the applicability to higher Reynolds number flows, it is scaled by a drag coefficient $C_d(Re_p, \eps_p)$.
Various such drag correlations are available \cite{BOGNER201571,tang_2015}.
The particle Reynolds number based on the superficial fluid velocity is
\begin{equation}
Re_p = \frac{\eps_{f|p} \rho_f d_p |\bs{u}_p-\bs{u}_{f|p}|}{\mu_0},
\end{equation}
with the local fluid volume fraction $\eps_{f|p} = 1-\eps_{p|p}$, the fluid density $\rho_f$ and the dynamic viscosity $\mu_0$.
Here, we make use of the drag correlation from \cite{TENNETI20111072} since it is derived as an extension to the single sphere case, $C_d(Re_p,0)$, and is valid for the regarded Reynolds numbers.

The pressure force, which features the gradient of the total fluid pressure, allows the particle to respond to local changes in the dynamic pressure. 
It also includes the buoyancy term if gravity is acting on the fluid in the simulation.
Otherwise, the buoyancy force $\bs{F}_{buoy} = -\rho_f V_p \bs{g}$ must be added to the particles in addition to the gravitational force $\bs{F}_g=\rho_p V_p \bs{g}$, with the gravitational acceleration $\bs{g}$.

The added, or virtual, mass force models the observation that an accelerating particle experiences a force against its acceleration since the fluid surrounding the particle has to be accelerated as well \cite{auton_hunt_prudhomme_1988,crowe2011multiphase}.
It thus appears as if the particle has gained, or added, mass.
Since in our simulations, the densities of the solid and the fluid phase are of the same order, we expect that the added mass force may have a non--negligible effect.

At last, also shear forces act on the particle and we account for them by including a shear--induced lift force \cite{saffman_1965}.

\subsubsection{Lubrication force}
\label{sec:lubrication}

When two immersed objects approach each other, the fluid between these objects gets squeezed out of the gap and this flow can produce a significant resistance to such a (wet) collision.
As a result, the rebound velocity after the collision is suppressed and a lower effective coefficient of restitution is observed.
Simulation models, that do not accurately resolve the flow inside these gaps, thus have to take care of the unresolved lubrication forces to obtain the correct rebound behavior.
In this work, we explicitly calculate the lubrication force $\bs{F}_{lub}$ and add it to the particle, Eq.~\eqref{eq:particle_motion}.
The lubrication force acting on a sphere $i$ due to another sphere $j$ can be computed as \cite{BALL1997444}:
\begin{equation}
\label{eq:lubrication_force}
\bs{F}_{lub,i} = -a_{sq} ( \bs{u}_{pr,ij} \cdot \bs{n}_{ij} ) \bs{n}_{ij} - a_{sh} \left(\frac{2}{h_{ij} + \langle d\rangle }\right)^2 [ \bs{u}_{pr,ij} - ( \bs{u}_{pr,ij} \cdot \bs{n}_{ij} ) \bs{n}_{ij}],
\end{equation}
with $\bs{n}_{ij}$ from Eq.~\eqref{eq:collision_normal}, the relative particle velocity $\bs{u}_{pr,ij} = \bs{u}_{p,i}-\bs{u}_{p,j}$, a mean diameter $\langle d\rangle=\tfrac{2d_{p,i}d_{p,j}}{d_{p,i} + d_{p,j}}$, and the gap size $h_{ij}$, calculated as
\begin{equation}
h_{ij} = |\bs{x}_{p,j}-\bs{x}_{p,i}| - \tfrac{1}{2}(d_{p,i} + d_{p,j}).
\end{equation}
The coefficients are given as
\begin{subequations}
	\begin{align}
	a_{sq} &= \frac{3}{2}\pi \mu_0 \langle d\rangle\left(\frac{\langle d\rangle}{4h_{ij}}+ \frac{18}{40}\ln\left(\frac{\langle d\rangle}{2h_{ij}}\right) + \frac{9}{84}\frac{h_{ij}}{\langle d\rangle}\ln\left(\frac{\langle d\rangle}{2h_{ij}}\right)\right),\\
	a_{sh} &= \frac{1}{2}\pi \mu_0 \langle d\rangle\ln\left(\frac{\langle d\rangle}{2h_{ij}}\right)\frac{(\langle d\rangle+h_{ij})^2}{4}.
	\end{align}
\end{subequations}
We currently do not include contributions from angular particle velocities and lubrication effects on the torque, which are also given in \cite{BALL1997444}.
For a sphere--wall interaction, we assume $d_{p,j}\rightarrow\infty$ and obtain $\langle d\rangle = 2d_{p,i}$.
We consider those forces only if the gap size is below a certain cut off value $h_c$.
Such a procedure is also commonly applied in fully resolved simulations \cite{nguyen_lub_2002}.
Additionally, we pose a lower limit on the gap size, such that $h_{ij} := \text{max}( h_{ij}, 10^{-5} \Delta x)$, to avoid too large forces \cite{BARTUSCHAT20151}.
The accuracy of this approach is demonstrated in Sec.~\ref{sec:sphere_wall_collision}. 

\subsection{Fluid phase}
\label{sec:fluid_phase}
Next, we present and discuss the models of the fluid flow where we account for the fluid volume displaced by the solid phase.
The corresponding LBM formulation is presented, as well as the evaluation of the fluid--particle interaction forces. 

\subsubsection{Generalized Navier--Stokes equations}
\label{sec:GNS}
The fluid flow inside a porous medium is commonly described via the so--called generalized Navier--Stokes (GNS) equations \cite{NITHIARASU19973955,guo_porous_2002}:
\begin{subequations}
	\label{eq:GNS}
	\begin{align}
	\nabla \cdot \bar{\bs{u}}_f &= 0, \label{eq:GNS_mass} \\
	\rho_f\left(\frac{\partial \bar{\bs{u}}_f}{\partial t} + (\bar{\bs{u}}_f \cdot \nabla) \left(\frac{\bar{\bs{u}}_f}{\eps_f}\right) \right) &= - \nabla(\eps_f p) + \mu_{e} \nabla^2\bar{\bs{u}}_f + \bs{f}_{p\rightarrow f} + \eps_f \bs{f}_b \label{eq:GNS_momentum},
	\end{align}
\end{subequations}
where $\eps_f$ is the fluid volume fraction that can vary locally, $\rho_f$ is the fluid density, $p$ the fluid pressure, $\bar{\bs{u}}_f = \eps_f \bs{u}_f$ is the volume--averaged fluid velocity, and $\mu_e$ is the effective fluid viscosity. 
For $\eps_f=1$, the incompressible Navier--Stokes equations are recovered.
The forcing acting on the fluid comprises two parts: 
The body force density $\bs{f}_b$ is induced by an external force.
The other part, $\bs{f}_{p\rightarrow f}$, accounts for the interphase momentum transfer due to the presence of the porous medium.
It usually consists of the Darcy and Forchheimer drag correlations which are the linear and nonlinear parts of Ergun's experimental drag correlation \cite{ergun_1952}.
In order to account for the velocity of the particle phase and to incorporate other important contributions, we enhance and modify this interphase momentum transfer term.

To define those quantities, a filter operation $\mathscr{D}$ must be introduced first that allows to distribute a particle property $\phi_p$ to the surrounding fluid cells in order to obtain the continuous field
\begin{equation}
\label{eq:distribution_filter}
\phi_f(\bs{x}) = \sum_{i=1}^{N_p} \mathscr{D}(\bs{x},\bs{x}_{p,i})\phi_{p,i}.
\end{equation}
In this work, we choose $\mathscr{D}$ to be the same as the previously defined interpolation filter $\mathscr{I}$ from Eq.~\eqref{eq:filter_def}.

With the help of Eq.~\eqref{eq:distribution_filter}, the fluid volume fraction in a cell can now be calculated from the particle positions via
\begin{equation}
\label{eq:fluid_volume_fraction}
	\eps_f(\bs{x}) = 1 - \eps_p(\bs{x}) = 1 - \frac{1}{(\Delta x)^3}\left(\sum_{i=1}^{N_p} \mathscr{D}(\bs{x},\bs{x}_{p,i}) V_{p,i}\right).
\end{equation}
In the same way, and with the modifications from Sec.~\ref{sec:particle_Ffp}, the force density accounting for the interphase momentum transfer is given as
\begin{equation}
\label{eq:forces_pf}
\bs{f}_\pf = - \frac{1}{(\Delta x)^3} \sum_{i=1}^{N_p}\mathscr{D}(\bs{x},\bs{x}_p)(\bs{F}_d + \bs{F}_l + \bs{F}_{am})_i.
\end{equation}
Note, that in contrast to the hydrodynamic force on the particle, Eq.~\eqref{eq:forces_fp}, we purposely do no include the pressure force $\bs{F}_{pr}$ here since this contribution is already present in the pressure term of the momentum GNS equation, Eq.~\eqref{eq:GNS_momentum}. 

Following \cite{finn_li_apte_2016}, a modified version of Eilers equation is used to include the effect of the solid phase on the dynamic viscosity $\mu_0$ \cite{Ferrini1979}:
\begin{equation}
\label{eq:visc_eilers}
\mu_{*} = \mu_0 \left(1+\frac{0.5[\mu]\eps_p}{(1-\eps_p)/\eps_{cp}}\right)^2,
\end{equation}
with the intrinsic viscosity $[\mu]=2.5$, that models the effect of the particle shape on the rheology of the fluid--particle system, and the solid volume fraction for a close sphere packing, $\eps_{cp}=0.64$.
Additionally, a large--eddy (LES) approach is applied to model the unresolved subgrid--scale turbulence which, for the simplest case of a Smagorinsky type turbulence model, adds a turbulent contribution $\mu_t$ to Eq.~\eqref{eq:visc_eilers} and will be discussed in more detail in the next section.
Thus, the effective viscosity $\mu_e$ is given as
\begin{equation}
\label{eq:visc_eff}
\mu_e = \mu_* + \mu_t.
\end{equation}

\subsubsection{Lattice Boltzmann formulation}
\label{sec:LBM}
For the simulation of the incompressible Navier--Stokes equations, we employ the lattice Boltzmann method \cite{chen_lattice_1998}.
It computes the evolution of particle distribution functions (PDFs) on a Cartesian lattice by solving the discretized version of the Boltzmann equation.
Each of these PDFs $f_q$ corresponds to a discrete lattice velocity $\bs{c}_q$.
A common choice is the \textit{D3Q19} lattice model \cite{qian_lattice_1992} that features a set of 19 lattice velocities, i.e. $q \in \{0,\dots,18\}$.
A lattice Boltzmann model for the generalized Navier--Stokes equations, Eqs.~\eqref{eq:GNS}, has been proposed by \cite{guo_porous_2002} and is given as
\begin{equation}
f_q( \bs{x} + \bs{c}_q\,\Delta t ,t+\Delta t) = f_q(\bs{x},t) + \tfrac{\Delta t}{\tau_e}\left(f_q^{\text{eq}}(\rho_f, \bar{\bs{u}}_f, \eps_f ) - f_q(\bs{x},t) \right) + \mathcal{F}_q(\bs{x},\bar{\bs{u}}_f,\eps_f,t)\Delta t.  \label{eq:LBM_Equation}
\end{equation}
It uses a single relaxation time parameter $\tau_e \in (\frac{1}{2}, \infty)$ to linearly relax the PDFs towards their equilibrium values $f_q^{\text{eq}}$, which can be computed via  
\begin{equation}
f_q^{\text{eq}}(\rho_f,\bar{\bs{u}}_f,\eps_f) = w_q \rho_f \left( 1 + \frac{\bs{c}_q \cdot \bar{\bs{u}}_f}{c_s^2} + \frac{(\bs{c}_q \cdot \bar{\bs{u}}_f)^2}{2\eps_fc_s^4} - \frac{\bar{\bs{u}}_f \cdot \bar{\bs{u}}_f}{2\eps_fc_s^2} \right). \label{eq:LBM_EQ}
\end{equation}
The lattice weights $w_q$ are as given e.g. in \cite{qian_lattice_1992} and $c_s$ is the lattice speed of sound.
The forcing operator in Eq.~\eqref{eq:LBM_Equation} is used to incorporate external forces and can be written as \cite{guo_porous_2002,guo_discrete_2002}:
\begin{equation}
\mathcal{F}_q(\bs{x},\bar{\bs{u}}_f,\eps_f,t) = w_q \left(1-\frac{\Delta t}{2\tau_e}\right) \left[\frac{\bs{c}_q}{c_s^2} -\frac{\bar{\bs{u}}_f}{\eps_fc_s^2} +  \frac{\bs{c}_q \cdot \bar{\bs{u}}_f}{\eps_fc_s^4} \bs{c}_q\right] \cdot \bs{f}^{\text{ext}}(\bs{x},t), \label{eq:Forcing}
\end{equation}
with a force density $\bs{f}^{\text{ext}} =\bs{f}_{p\rightarrow f} + \eps_f \bs{f}_b$.
The fluid density $\rho_f$ and the volume--averaged fluid velocity $\bar{\bs{u}}_f$ are cell local quantities and calculated via moments of the PDFs:
\begin{align}
\rho_f(\bs{x},t) & = \sum_q f_q(\bs{x},t),\\
\bar{\bs{u}}_f(\bs{x},t) &= \frac{1}{\rho_f}\sum_q f_q(\bs{x},t) \bs{c}_q + \frac{\Delta t}{2\rho_f} \bs{f}^{\text{ext}}(\bs{x},t).\label{eq:MacVel}
\end{align}
The pressure is given as $p = c_s^2 \rho_f / \eps_f$. 
The effective dynamic viscosity $\mu_{e}$ is related to the relaxation time $\tau_e$ via 
\begin{equation}
\label{eq:visc_tau}
\mu_{e} /\rho_f = (\tau_e - \tfrac{\Delta t}{2})c_s^2.
\end{equation}
Analogously to Eq.~\eqref{eq:visc_eff}, this relaxation time includes two contributions, 
\begin{equation}
	\label{eq:tau_e}
	\tau_e = \tau_* + \tau_t,
\end{equation}
where $\tau_*$ and $\tau_t$ are the relaxation times that correspond to the modified fluid viscosity $\mu_*$ from Eq.~\eqref{eq:visc_eilers} and the turbulence viscosity $\mu_t$, respectively \cite{hou1996lattice,YU2005599}:
\begin{subequations}
	\begin{align}
	\mu_* / \rho_f & = \nu_* = (\tau_* - \tfrac{\Delta t}{2}) c_s^2, \label{eq:visc_tau_*}\\
	\mu_t/\rho_f &= \nu_t = \tau_t c_s^2 
	\end{align}
\end{subequations}
For the Smagorinsky type LES turbulence model, the turbulence viscosity is obtained from the strain rate tensor $S = S_{\alpha,\beta} = \tfrac{1}{2}( (\nabla \bar{\bs{u}}_f) + (\nabla\bar{\bs{u}}_f)^\top) $ as \cite{hou1996lattice}:
\begin{equation}
\label{eq:visc_turb}
\nu_t = (C_s \Delta x_{\text{LES}})^2 \bar{S}, \quad \bar{S} = \frac{\bar{Q}}{2\rho_fc_s^2 \tau_e},\quad \bar{Q} = \sqrt{2 \sum_{\alpha, \beta} \bar{Q}_{\alpha,\beta}\bar{Q}_{\alpha,\beta}},
\end{equation}
with the Smagorinsky constant $C_s = 0.1$ \cite{YU2005599}, a filter length $\Delta x_{\text{LES}} = \Delta x$, and the (filtered) mean momentum flux $\bar{Q}$.
The latter one is computed from the momentum fluxes $Q_{\alpha,\beta}$, obtained as the second--order moments of the nonequilibrium parts of the distribution functions \cite{guo_porous_2002,hou1996lattice}:
\begin{equation}
Q_{\alpha,\beta} = \sum_q c_{q,\alpha} c_{q,\beta}[f_q - f_q^{\text{eq}}].
\end{equation}
This leads to a quadratic equation for $\tau_t = \nu_t / c_s^2$, with the solution given as \cite{hou1996lattice}   
\begin{equation}
\label{eq:tau_t}
\tau_t = \frac{1}{2}\left(\sqrt{\tau_*^2 + 2\sqrt{2}(C_S\Delta x_{\text{LES}})^2(\rho_fc_s^4)^{-1}\bar{Q}} - \tau_* \right).
\end{equation}

In the context of the LBM, all quantities are commonly expressed in terms of lattice units which results in $\Delta t = 1$, $\Delta x = 1$, $c_s = \tfrac{1}{\sqrt{3}}$, and $\rho_f\approx 1$.

\subsubsection{Fluid--particle interaction forces}
\label{sec:lbm_interaction_force}

The components of the fluid--particle interaction force density $\bs{f}_\pf$ are given in Eq.~\eqref{eq:forces_pf} and are included via $\bs{f}^\text{ext}$ in Eq.~\eqref{eq:Forcing}.
Their definitions are as discussed in Sec.~\ref{sec:particle_Ffp}.
Their evaluation requires macroscopic fluid quantities and their derivatives instead of the readily available LBM quantities, i.e.~the particle distribution functions. 
We therefore briefly discuss the LBM specific aspects of the force evaluation here.

In general, we compute and store all required quantities explicitly for each computational cell in the simulation domain.
Those can then be interpolated to the particle position via Eq.~\eqref{eq:interpolation}.
The velocity used to compute the interaction forces is the fluid--phase velocity $\bs{u}_f$ rather than volume--averaged fluid velocity.
To evaluate the appearing derivatives, we use the lattice differential operators from \cite{ramadugu_2013_lattice_operators}.
Consequently, the gradient of a scalar quantity $\phi$, as e.g. $P = \eps_f p $ in the pressure gradient force, can be approximated as
\begin{equation}
\nabla \phi \approx \frac{1}{c_s^2} \sum_q w_q \bs{c}_q \phi(\bs{x}+\bs{c}_q\Delta t).
\end{equation}
The rotation of a vector field $\bs{\psi}$, as e.g. the velocity field in the lift force, is computed as
\begin{equation}
\nabla \times \bs{\psi} \approx \frac{1}{c_s^2} \sum_q w_q \bs{c}_q \times \bs{\psi}(\bs{x}+\bs{c}_q\Delta t).
\end{equation}

A factor that significantly influences the stability of this specific fluid--particle coupling approach originates from the definition of the macroscopic velocity in Eq.~\eqref{eq:MacVel}.
When the external force is changed, e.g. by adding the interaction force, the macroscopic fluid velocity changes also.
As a consequence, a reevaluation of the interaction force components, that depend on the velocity, with the updated velocity values will yield distinct force values.
If not handled correctly, this can easily lead to severe oscillations of the force term which then affect the overall stability of the simulation.
A cell--local implicit treatment of this dependency for each component of the interaction force would be desirable, as e.g. done in \cite{guo_porous_2002} for Ergun's drag correlation.
However, this is not applicable here since in general the applied (drag) correlations are more complex than Ergun's correlation and the influence of all present particles would have to be treated implicitly at the same time.
We therefore rely on an iterative reevaluation of the required fluid quantities with the updated force contributions.
This procedure works well for e.g.~the drag force which is usually the dominant contribution.
In our simulations, we observe that small oscillations in the drag force in combination with the added mass force, which depends on the temporal change of the fluid velocity, are often the source of growing instabilities, especially for large solid volume fractions.
In order to damp this feedback mechanism, we refrain from reevaluating $\tfrac{\text{D}\bs{u}_f}{\text{D}t}$, see Tab.~\ref{tab:InteractionForces}, during those internal iterations.
Since $\tfrac{\text{d}\bs{u}_p}{\text{d}t}$ is updated in each iteration, the added mass force still changes.
For additional stability, especially in cases of large solid volume fractions, we reevaluate the fluid velocity after the contributions from the added mass and lift forces have been computed and before the drag force is evaluated.  

\subsection{Complete DPS algorithm for coupled LBM--DEM simulations}
\begin{algorithm}[t]
	Initially, evaluate and store fluid volume fraction $\eps_f$ in each cell, Eq.~\eqref{eq:fluid_volume_fraction}\;
	Apply external body force, $\eps_f \bs{f}_b$, to cells\;
	\For{each time step}{
		Evaluate and store $\tfrac{\text{D}\bs{u}_f}{\text{D}t}$\;
		\For{each interaction subcycle}{
			Evaluate and store $\bs{u}_f$, $\nabla P$, $\nabla \times \bs{u}_f$\;
			\For{each particle}{
				Evaluate $\bs{F}_l$ and distribute to cells\;
				Evaluate $\bs{F}_{am}$ and distribute to cells\;
			}
			Reevaluate $\bs{u}_f$\;
			\For{each particle}{	
				Evaluate $\bs{F}_{pr}$\;
				Evaluate $\bs{F}_d$ and distribute to cells\;
				Apply gravitational force $\bs{F}_g$\;
			}
			\For{each DEM substep}{
				Compute lubrication forces $\bs{F}_{lub}$ on all particles, Eq.~\eqref{eq:lubrication_force}\;
				Compute collision forces $\bs{F}_c$, Eq.~\eqref{eq:forces_collision}, and torques, $\bs{T}_c$, Eq.~\eqref{eq:torques_collision}\;
				Perform time integration of particles, Eqs.~\eqref{eq:particle_motion}, with time step size $\Delta t_\text{DEM}$\;
			}
			Reevaluate fluid volume fraction $\eps_f$\;
		}
		Compute $\tau_*$, Eq.~\eqref{eq:visc_tau_*}, from modified viscosity $\mu_*$, Eq.~\eqref{eq:visc_eilers}\;
		Compute $\tau_t$, Eq.~\eqref{eq:tau_t}, and then $\tau_e$, Eq.~\eqref{eq:tau_e}\; 
		Perform GNS--LBM step, Eq.~\eqref{eq:LBM_Equation}, with time step size $\Delta t$\;
	}
	\caption{Overview of the LBM--DEM DPS algorithm.}
	\label{alg:DPS}
\end{algorithm}
\label{sec:DPSalgo}
A detailed overview of our proposed algorithm to carry out discrete particle simulations by coupling the lattice Boltzmann method and the discrete element method is presented in Alg.~\ref{alg:DPS}.
It features an outer time loop with time step size $\Delta t$ which is also the time step used in the LBM simulations.
However, the typical time scales of particle collisions are usually significantly smaller than the time scales of the fluid solver.  
Therefore, the time loop is subdivided twice, such that during one time step $n_\text{int}$ interaction subcycles are performed which include $n_\text{DEM}$ DEM substeps.
The ratio between the LBM and the DEM time step is then $\Delta t / \Delta t_\text{DEM} = n_\text{int} n_\text{DEM}$ and easily reaches $\mathcal{O}(100)$, see Sec.~\ref{sec:sphere_wall_collision}.

The interaction subcycles are introduced to improve the accuracy of the interaction force evaluations since the particles' position, needed for the interpolation of fluid quantities, as well as their velocity are updated in each subcycle.
Additionally, this reevaluation of the interaction forces helps to avoid oscillations in the force components effectively, see the discussion in Sec.~\ref{sec:lbm_interaction_force}.

The additional DEM substeps are used to deal efficiently with simulation scenarios that require accurate resolution of the collision and lubrication interactions between particles, as e.g.~in Sec.~\ref{sec:sphere_wall_collision}, or flow through densely packed particle beds.
During those substeps, the currently acting forces on the particles, $\bs{F}_\fp$ and $\bs{F}_g$, are kept constant, and only collision and lubrication contributions are updated.
Since the computationally expensive evaluation of the interaction forces is thus omitted, the usage of DEM substeps is more efficient than increasing the number of interaction subcycles.

Finally, we note that no global data exchange is required in our algorithm at any time since all parts operate only on locally available data.
This key feature ensures its efficient applicability to large--scale simulations on HPC systems. 

\section{Hindered settling behavior of spheres}
\label{sec:hinderedSettling}

In order to reliably capture the fluid--solid interaction in various applications, the DPS approach must predict the correct behavior over a wide range of solid volume fractions, i.e. from single particles to densely packed scenarios.
A suitable test setup to validate these cases is the evaluation of the hindered settling behavior of spheres at various solid volume fractions.
It is well known that the average relative settling velocity of the spheres $\langle u_{pS}\rangle$ in an unbounded domain is affected by the average solid volume fraction $\langle\eps_p\rangle$ as follows \cite{RICHARDSON195465}:
\begin{equation}
\label{eq:hinderedSettlingCorrelation}
\frac{\langle u_{pS}\rangle}{u_{pT}} = (1-\langle\eps_p\rangle)^\kappa,
\end{equation}
where $u_{pT}$ is the terminal settling velocity of a single sphere in an infinite fluid, which can be measured in experiments or found by direct numerical simulations.
The coefficient $\kappa$ is a function of the Reynolds number $Re_T = \rho_fu_{pT}d_p/\mu_0$ and its value for unbounded flows can be estimated by the widely used semi--empirical correlations from Richardson and Zaki \cite{RICHARDSON195465}:
\begin{equation}
\label{eq:hinderedCoefficientRZ}
\kappa = 
\begin{cases}
4.65, & Re_T < 0.2 \\
4.35\, Re_T^{-0.03}, & 0.2 \leq Re_T < 1 \\
4.45\, Re_T^{-0.1}, & 1 \leq Re_T < 500 \\
2.39, & 500 \leq Re_T \\
\end{cases}
\end{equation}

\subsection{Setup}
We use a fully periodic domain of size $[L_x \times L_y \times L_z]/d_p = 32 \times 32 \times 32$.
Initially, spheres with a uniform diameter $d_p$ are randomly generated inside this domain until the desired average solid volume fraction is obtained.
In order to have non--overlapping spheres before starting the coupled simulations, several DEM time steps are carried out to resolve possible overlaps.
Then, their velocity is set to zero.
These spheres at rest then begin to settle under gravity in the viscous fluid.
Regarding the physical parameters, we follow the experimental setup from \cite{BALDOCK200491} which is also used in simulations of \cite{finn_li_apte_2016}.
Thus, the fluid phase is water with $\rho_f = 1000\,$kg/m$^3$ and $\mu_0 = 0.001\,$Pa$\,$s.
The glass spheres have the properties $d_p = 0.35\cdot 10^{-3}\,$m, $\rho_p = 2500\,$kg/m$^3$, $e_n=0.88$, and $\mu_c = 0.25$.
Furthermore, we use $T_c = \Delta t/2$, $h_c=d_p$, $n_\text{int}=10$, and $n_\text{DEM}=50$.
The gravitational acceleration is $g = -9.81\,$m/s$^2$ and acts only on the spheres.
A uniform body force density $\bs{f}_b = (0,0,-\langle\eps_p \rangle g ( \rho_p - \rho_f))^\top$ is applied on the fluid to balance the spheres' weight.
In the experiments of \cite{BALDOCK200491}, the terminal settling velocity of a single sphere $u_{pT}= -0.048\,$m/s was measured, which results in $Re_T = 16.8$.
The average relative sphere velocity in gravitational direction is evaluated as
\begin{equation}
\langle u_{pS}\rangle = \frac{1}{N_p}\sum_{i=1}^{N_p} u_{p,i} - \frac{(\Delta x)^3}{L_xL_yL_z}\sum_{\bs{x}} u_f(\bs{x})
\end{equation}
The simulations are run for $30t_{St}$ time steps, with $t_{St} = \rho_p d_p^2/ 18 \mu_0$, which is long enough to obtain a stationary average settling velocity \cite{finn_li_apte_2016}. 
The number of spheres ranges from $1$ to $37549$, corresponding to $\langle \eps_p \rangle \approx 0$ up to $0.6$.
We use three different sphere diameters, $d_p/\Delta x = \tfrac{1}{4},\tfrac{1}{2},$ and $1$, to investigate resolution effects. 

\subsection{Results}

\begin{figure}[t]
	\centering
	\includegraphics[width=0.5\textwidth]{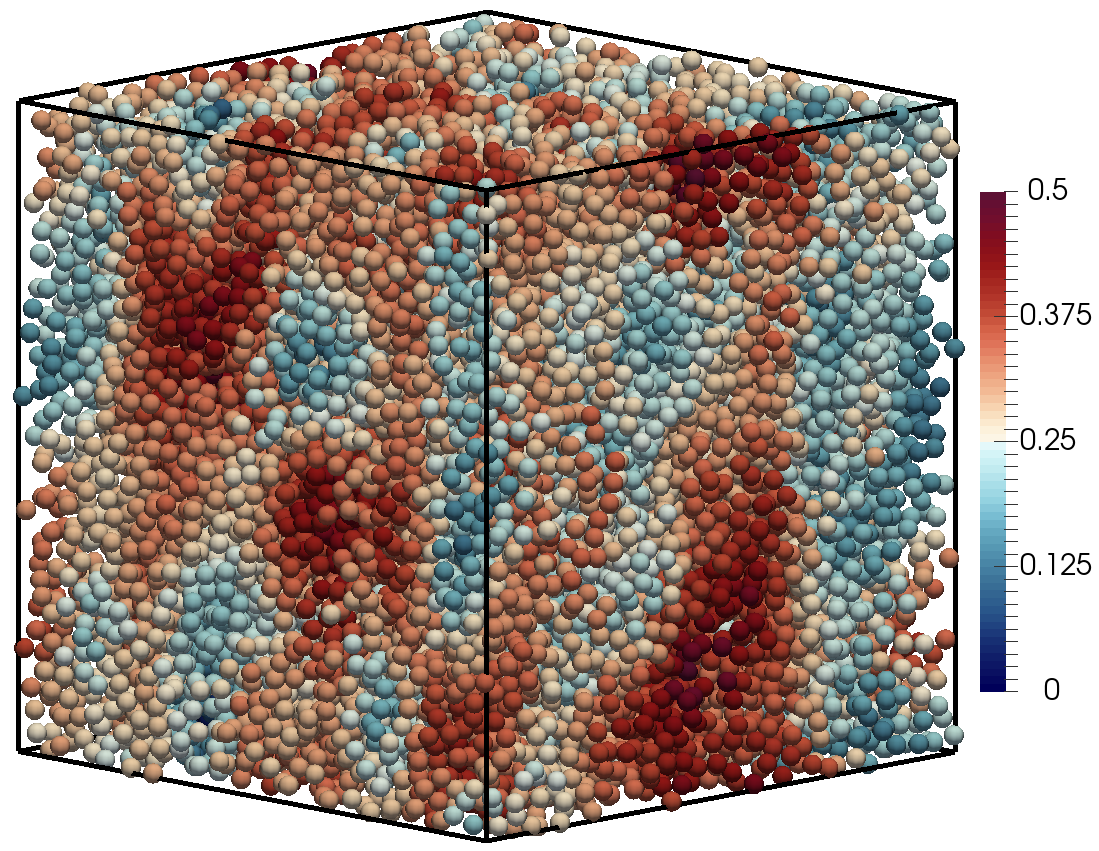}
	\caption{Visualization of the setup for $\langle \eps_p \rangle = 0.3$ and $d_p=0.5$. The color indicates the normalized relative settling velocity of the individual spheres, $u_{pS}/u_{pT}$. Gravity acts in vertical direction.}
	\label{fig:hinderedSettlingVis}
\end{figure}

\begin{figure}[t]
	\centering
	\input{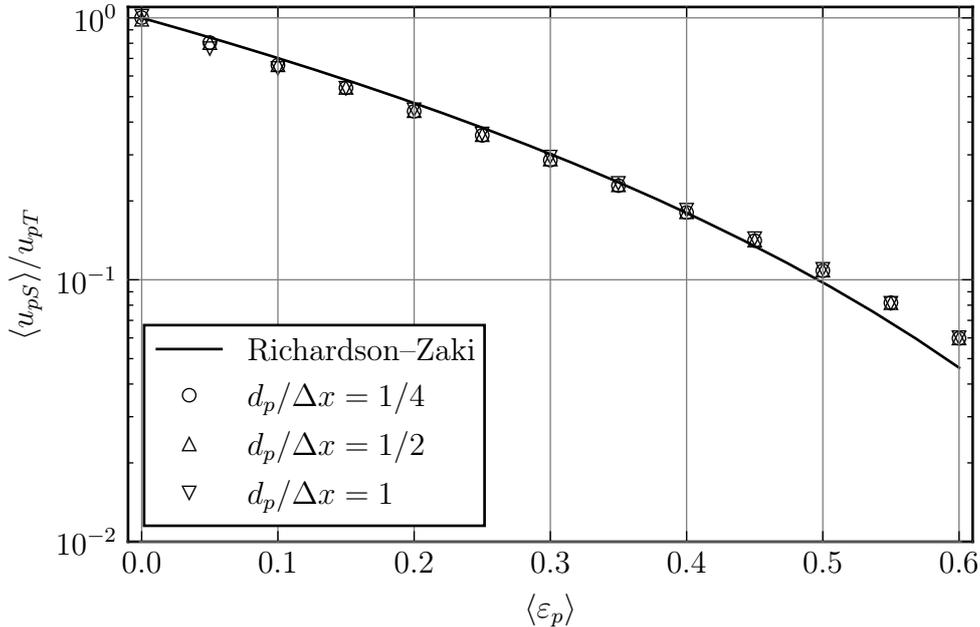}
	\caption{Average relative settling velocity normalized by the single sphere terminal settling velocity as a function of the average solid volume fraction $\langle\eps_p\rangle$. For comparison, the Richardson--Zaki correlation \cite{RICHARDSON195465}, Eqs.~\eqref{eq:hinderedSettlingCorrelation} \& \eqref{eq:hinderedCoefficientRZ}, is included.}
	\label{fig:hinderedSettling}
\end{figure}

A visualization of the outcome of this test case is shown in Fig.~\ref{fig:hinderedSettlingVis}.
As can be seen from the spheres' velocities, clusters of similar velocity form during the simulation.
The results of the study are plotted in Fig.~\ref{fig:hinderedSettling} with the average relative settling velocity over the average solid volume fraction.
Apparently, the simulated settling velocities match the predictions from the Richardson--Zaki formula very well over the whole range of solid volume fractions.
In particular, the settling behavior of a single sphere is captured well.
For the densely packed scenarios slightly larger settling velocities are obtained.
Altering the ratio between sphere diameter and cell size does not change the settling velocity.

\subsection{Sensitivity analysis}

\begin{table}[t]
	\centering
	\begin{tabular}{c|ccccc|c}
		case & lift force & lubrication force & turbulence model & mod. viscosity & $n_\text{int}$ & $\langle u_{pS}\rangle/u_{pT}$\\\hline
		A0 & \checkmark & \checkmark & \checkmark & \checkmark & 10 & 0.288 \\\hline
		A1 & - & \checkmark & \checkmark & \checkmark & 10 & 0.297 \\
		A2 & \checkmark & - & \checkmark & \checkmark & 10 & 0.288 \\
		A3 & \checkmark & \checkmark & - & \checkmark & 10 & 0.288 \\
		A4 & \checkmark & \checkmark & \checkmark & - & 10 & 0.289 \\
		A5 & \checkmark& \checkmark & \checkmark & \checkmark & 1 & 0.288 \\
	\end{tabular}
	\caption{Simulation parameters for the sensitivity study and their influence on the settling velocity for $\langle \eps_p \rangle = 0.3$ and $d_p/\Delta x = \tfrac{1}{2}$. A0 is the baseline case.}
	\label{tab:hinderedSensitivity}
\end{table}

Next, we focus on a specific scenario and investigate the influence of different model parameters on the obtained settling velocity.
With such a sensitivity analysis, the most relevant parameters can be identified.
This is essential to develop computationally efficient numerical approaches since contributions with small influence can possibly be neglected.
As a setup, we chose $d_p/ \Delta x=\tfrac{1}{2}$ and $\langle \eps_p \rangle = 0.3$, as visualized in Fig.~\ref{fig:hinderedSettlingVis}.
The parameters of the cases under comparison are as summarized in Tab.~\ref{tab:hinderedSensitivity}.
We evaluate the contributions of the lift force, lubrication forces, the turbulence model, the modified viscosity of Eq.~\eqref{eq:visc_eilers}, and the number of interaction subcycles.
Without the lift force (A1), the settling velocity increases by around $3\%$ in comparison to the baseline case A0.
Lubrication forces are seemingly not important in this rather homogeneous setup since the settling velocity remains the same when neglected (A2).
Also the influence of the turbulence model is negligible here as the Reynolds number is not large enough such that turbulent flow structures would become dominant (A3).
The modifications of the fluid viscosity introduced by Eq.~\eqref{eq:visc_eilers} also do not alter the settling velocity significantly (A4).
Finally, decreasing the number of interaction subcycles yields the same settling velocity (A5), but the overlaps between the spheres become around four times larger since also $\Delta t_\text{DEM}$ is reduced. 

\subsection{Discussion}

The results of this test case show that our approach is capable of correctly predicting the fluid--particle interactions over a wide range of solid volume fractions, from a single sphere up to systems that are close to the maximum packing fraction.
The drag force, and thus the applied drag correlation, is the most important factor in this setup but also the volume exclusion effects on the fluid phase become relevant for larger solid volume fractions.
Even though the sensitivity study showed no changes for a smaller number of interaction subcycles, scenarios with large solid volume fractions require more than one internal iteration to yield stable results for the reasons mentioned in Sec.~\ref{sec:lbm_interaction_force}.
We expect that the influence from using the modified viscosity or the turbulence model will become more important in other setups like a shear flow over a sediment bed at high Reynolds numbers which also features large velocity gradients.
In such a case, also lift forces will play a more prominent role on the sediment motion.
The observation that the simulated settling velocities are independent of the applied sphere diameter without tuning other numerical parameters like the relaxation time, necessary e.g.~in the PIBM approach from \cite{ZHANG20151}, again demonstrates the validity of our approach.
This property offers a great amount of flexibility as the ratio between diameter and cell size can thus be chosen with respect to the requirements of the application setup.
It also enables the accurate simulation of polydisperse systems, i.e.~mixtures of small and large particles.
Furthermore, the approach can be combined with grid refinement techniques where the particles move across meshes with different grid sizes $\Delta x$.

\section{Sphere--wall collision in a viscous fluid}
\label{sec:sphere_wall_collision}

In this scenario, a sphere, heavier than the surrounding viscous fluid, is dropped in a box, and its trajectory and velocity are tracked.
When it hits the bottom plane, a rebound will happen which allows to define an effective, or wet, coefficient of restitution
\begin{equation}
\label{eq:coeff_restitution_wet}
e_n^{wet} = -\frac{u_{pR}}{u_{pT}},
\end{equation}
where $u_{pR}$ is the rebound velocity of the sphere and $u_{pT}$ the terminal settling velocity.
From experiments \cite{joseph_zenit_hunt_rosenwinkel_2001,gondret_2002}, it is known that the sphere motion, and thus the ratio between the wet and dry coefficient of restitution, is mainly a function of the Stokes number,
\begin{equation}
St = \frac{\rho_p}{\rho_f} \frac{Re_T}{9},
\end{equation}
with $Re_T = \rho_f u_{pT}d_p / \mu_0$.
With this setup, the accuracy and correctness of the collision and lubrication forces as well as the substepping procedure will be tested.

\subsection{Setup}

In our simulations, we use a rectangular domain of size $[L_x\times L_y \times L_z]/d_p = 32 \times 32 \times 512$ and initialize a resting sphere horizontally centered and close to the upper boundary.
The domain is large enough such that the terminal velocity is always reached before the sphere impacts on the bottom wall.
For the fluid phase, no--slip boundary conditions are applied in all directions.
The gravity, only acting on the sphere, is given by a gravitational acceleration $g=-0.0001$ in lattice units.
The dry coefficient of restitution is $e_n = 0.97$ and the coefficient of friction is $\mu_c = 0.1$, which corresponds to a steel sphere and is similar to the one used in the experiments of \cite{gondret_2002}.
In order to keep the maximum penetration depth below $5\%$ of the sphere diameter, we use $T_c = \frac{d_p}{\Delta x}\Delta t $, see Eq.~\eqref{eq:collision_time}.
We keep the number of interaction subcycles at $n_\text{int}=10$ and choose $n_\text{DEM}=25\frac{\Delta x}{d_p}$ to maintain a constant number of DEM steps per $T_c$.
The cut--off gap size for the lubrication forces is $h_c = d_p$.
We define the impact time $t_I$ as the instance in time where the sphere loses contact with the wall, i.e. the collision force becomes zero after the first collision.
As in \cite{kidanemariam_uhlmann_2014}, various combinations of density ratios and fluid viscosities are employed in order to obtain different Stokes numbers $St \in [10, 2000]$.
Additionally, three different sphere diameters are tested, $d_p / \Delta x = \tfrac{1}{4}, \tfrac{1}{2}$, and $1$.

\subsection{Results}

\begin{figure}[t]
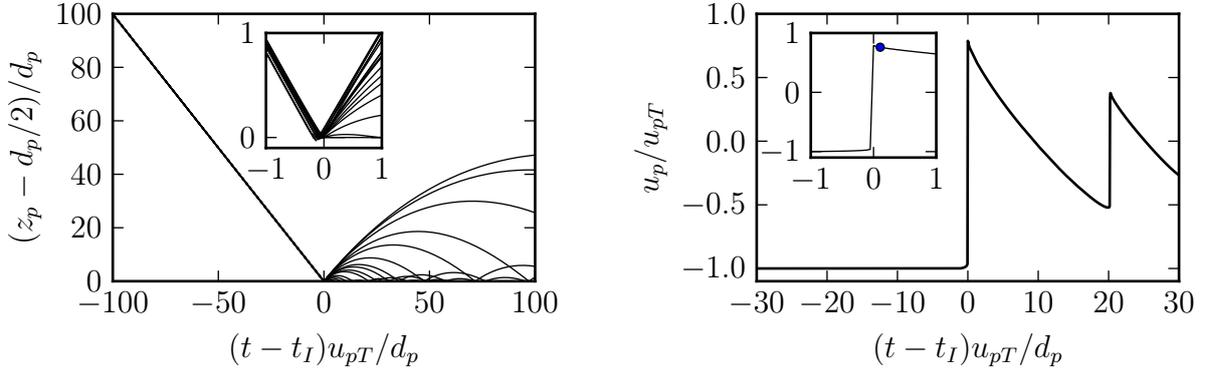

	\begin{subfigure}[b]{0.5\textwidth}
		\centering
		\input{figures/bouncingTrajectory.pgf}
	\end{subfigure}
	\begin{subfigure}[b]{0.5\textwidth}
		\centering
		\input{figures/velocityTrajectory.pgf}
	\end{subfigure}
	\caption{Left: Wall--normal distance of the sphere surface to the wall over time, shifted such that the times of the first impact coincide, for all simulated $St$ and $d_p/\Delta x = \tfrac{1}{2}$. Right: Normalized sphere velocity of $St=128$ over time, again shifted. Inset with marker of the rebound velocity evaluated at $t_I+t_R$.}
	\label{fig:bouncingTrajectories}
\end{figure}

In the left plot of Fig.~\ref{fig:bouncingTrajectories}, the trajectories of all simulations with $d_p / \Delta x= \tfrac{1}{2}$ are shown as wall--normal distances to the bottom wall as a function of time, which is shifted by the impact time and normalized with the reference time $d_p/u_{pT}$.
As can be seen from the overlapping, straight lines before the impact, all spheres obtain their terminal settling velocity before any interaction with the wall happens.
The zoomed inset shows that in some cases the distance gets negative, indicating that overlaps between the sphere and the wall happen which, however, are very small.
The maximum rebound height ranges from negligible small values for $St$ below $17$, which can be quantified as no rebound, up to around $48d_p$ for the largest simulated $St$.
The right plot depicts the temporal evolution of the sphere velocity for $St=128$. 
The inset of this plot shows that already a few time steps before the actual collision with the wall, the sphere gets decelerated due to lubrication forces. 
As marked in the inset, the rebound velocity is not taken as the maximum velocity after the collision but evaluated some time $t_R = 0.1 d_p/ u_{pT}$ after the impact.
This has been proposed by \cite{kidanemariam_uhlmann_2014}, to account for the temporal resolution of experimental measurement facilities used by \cite{joseph_zenit_hunt_rosenwinkel_2001} and \cite{gondret_2002}, to which the results are compared next.

\begin{figure}[t]
	\centering
	\input{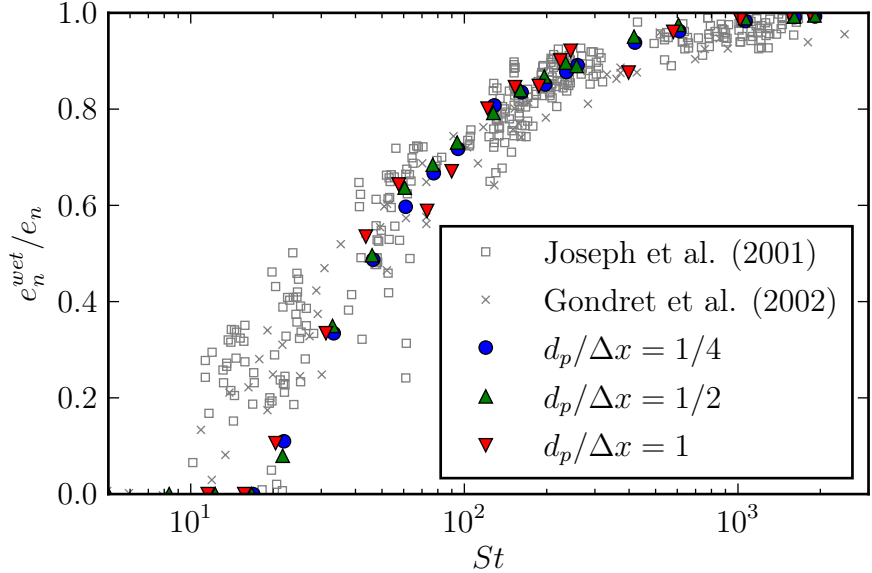}
	\caption{Ratio between the wet and dry coefficient of restitution as a function of the Stokes number $St$. For comparison, experimental data of \cite{joseph_zenit_hunt_rosenwinkel_2001} and \cite{gondret_2002} are included.}
	\label{fig:bouncingCoeff}
\end{figure}

The normalized wet coefficient of restitution as a function of $St$ is evaluated according to Eq.~\eqref{eq:coeff_restitution_wet} and compared to experiments, all shown in Fig.~\ref{fig:bouncingCoeff}.
It can be seen that the simulation results fit to the experimental data very well over the whole range of $St$ and also show an exponential increase.
In particular, the critical Stokes number, below which no rebound can be seen, agrees with the experimentally reported ones.
For $St>600$, the wet coefficient of restitution approaches the dry one which is also in accordance with the reported values of this so--called elastic limit.
The overall behavior is independent of the particle diameter used. 
Some perturbations can be observed, however, especially for $d_p/\Delta x = 1$, which can be explained by the low temporal resolution applied for evaluating the rebound velocity since $t_R$ has typically a magnitude of a few simulation time steps.

\subsection{Sensitivity analysis}
\label{sec:bounce_sensitivity}

\begin{table}[t]
	\centering
	\begin{tabular}{c|ccc|c}
		case & $n_\text{DEM}$ & added mass & $h_c$ & $ e_n^{wet}/e_n$ \\\hline
		B0 & 50 & \checkmark & $d_p$ & $0.792$ \\\hline
		B1 & 5 & \checkmark & $d_p$ & $0.899$ \\
		B2 & 500 & \checkmark & $d_p$ & $0.803$ \\
		B3 & 50 & - & $d_p$ &  $0.831$\\
		B4 & 50 & \checkmark & 0 & $0.976$ \\
		B5 & 50 & \checkmark & $d_p$/2 & $0.797$ \\
	\end{tabular}
	\caption{Simulation parameters for the sensitivity study and their influence on the ratio between wet and dry coefficient of restitution for $St=128$ and $d_p/ \Delta x = \tfrac{1}{2}$. B0 is the baseline case.}
	\label{tab:bouncingSensitivity}
\end{table}

\begin{figure}[t]
	\centering
	\input{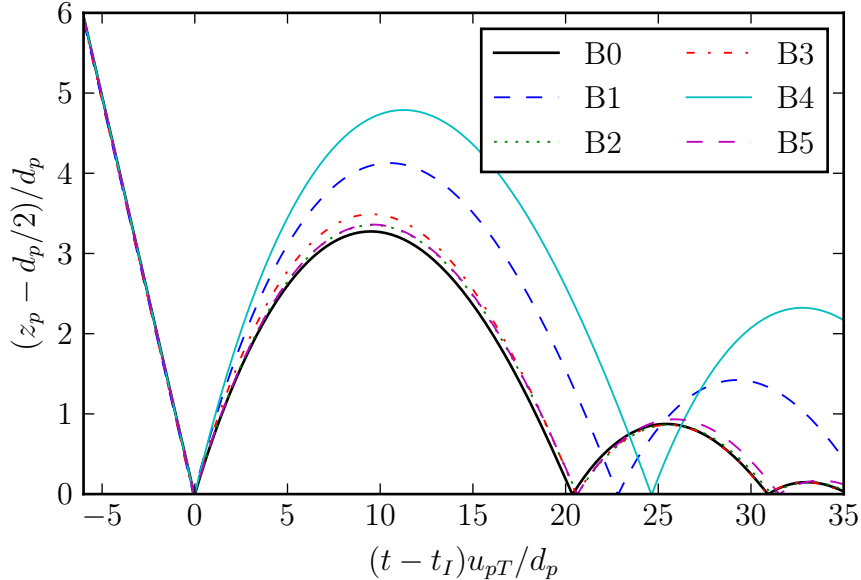}
	\caption{Wall--normal distance of the sphere surface to the wall over time for the setups of the sensitivity study, given in Tab.~\ref{tab:bouncingSensitivity}, for $St=128$.}
	\label{fig:bouncingSensitivityTrajectories}
\end{figure}

We again carry out a sensitivity analysis for this setup to quantify the influence of different numerical and model parameters.
We here investigate the effect on the relative coefficient of restitution and the rebound trajectory for the scenario with $St=128$ and $d_p/ \Delta x = \tfrac{1}{2}$.
These parameters are $n_\text{DEM}$, and consequently $\Delta t_\text{DEM}$, the added mass force contribution, and the cut off distance $h_c$ in the lubrication forces, with values given in Tab.~\ref{tab:bouncingSensitivity}.
The resulting trajectories are shown in Fig.~\ref{fig:bouncingSensitivityTrajectories}.
When decreasing $n_\text{DEM}$, and thus increasing $\Delta t_\text{DEM}$, the maximum rebound height as well as $e_n^{wet}$ increases since the collision and lubrication interaction is less accurately resolved.
Increasing $n_\text{DEM}$ hardly changes the outcome in comparison to the baseline case, indicating that $n_\text{DEM}=50$ is a good choice in this setup in terms of accuracy and computational efficiency.
Neglecting the added mass contributions to the interaction force (B3) slightly increases the rebound height and the coefficient of restitution.
The largest influence has the lubrication force since the rebound behavior is completely changed when no lubrication forces are considered (B4).
The value of the wet coefficient of restitution approaches the dry one and the rebound is significantly higher. 
As (B5) shows, the results vary only slightly when choosing a smaller cut--off distance than in the baseline case. 

\subsection{Discussion}

With the accuracy of the simulated settling velocity demonstrated in Sec.~\ref{sec:hinderedSettling}, this benchmark demonstrates that also single collision events in a submerged setup are well predicted with our approach.
In such a scenario, a fine enough temporal resolution of the collision dynamics has to be ensured to prevent large overlaps during the collision.
Furthermore, depending on fluid flow properties, the inclusion of lubrication forces can be crucial for a correct behavior.
In our approach, where we explicitly compute and set the lubrication forces for each sphere--sphere or sphere--wall pair, this evaluation can become the computationally most expensive part of the simulation in systems with a large number of particles. 
Therefore, an other approach could be used where the coefficient of restitution is changed dynamically for each collision, like in \cite{finn_li_apte_2016}.
This requires to compute a local impact Stokes number and to explicitly prescribe how the wet coefficient of restitution changes depending on this number.
However, such an approach introduces further modeling assumptions and is unable to capture physical effects like the deceleration of the sphere already before the collision happens.
Besides the lubrication and collision forces, also added mass forces can influence the dynamics of the system and should thus be included.

\section{Conclusion}
\label{sec:conclusion}

We have presented a novel algorithm for discrete particle simulations that couples the lattice Boltzmann method with the discrete element method.
The main feature of such simulations is that the solid phase consists of Lagrangian particles that are generally smaller than a computational cell used in the fluid simulation.
The coupling was realized by evaluating interphase momentum transfer forces which include drag, pressure gradient, lift, and added mass forces.
Here, special care and measures had to be taken inside the algorithm to prevent destabilizing oscillations.
The DEM was used to model the collision between the particles, accounting for normal and tangential collision forces.
For the description of the fluid flow, a LBM formulation of the generalized Navier--Stokes equations was chosen to include volume displacement effects due to the presence of the particulate phase.
Furthermore, we used a modified fluid viscosity based on the local fluid volume fraction and a Smagorinsky--type LES turbulence model.
The simulation results of our approach agreed well with semi--empirical predictions for the settling velocity of spheres over a large range of solid volume fractions, from a single sphere up to densely packed scenarios.
Additionally, we explicitly computed interparticle lubrication forces to obtain the correct collision behavior for submerged particles.

The major benefit of such a DPS approach in comparison to fully resolved simulations is that significantly more particles can be simulated in less time. 
This enables efficient simulations of larger systems like fluidized beds or riverbeds as they are relevant in industrial and engineering applications.
In order to obtain predictive results with the DPS technique, a careful validation of the method is essential.
For our approach, we demonstrated accurate simulation results for different configurations, demonstrating their suitability for more complex scenarios.
This approach also offers the flexibility to be further extended by exchanging the applied models and adding further ones.
For example, even though we focused on spherical particles in this work, the presented approach can be extended to other shapes, which then allows for a more realistic representation of e.g. river sediments, as done in \cite{finn_li_apte_2016}.
Also, this approach can be used in combination with a fully resolved simulation approach such that polydisperse systems with large particle size ratios can be simulated.
Then, the large particles could be geometrically fully resolved and the smaller particle are treated by DPS.

\bibliographystyle{elsarticle-num}
\bibliography{Library}	
\end{document}